# Controlling structural distortion in the geometrically frustrated layered cobaltate YBaCo$_4$O$_{7+\delta}$ by Fe substitution and its role on magnetic correlations

A K Bera[†], S M Yusuf*, and S S Meena

*Solid State Physics Division, Bhabha Atomic Research Centre, Mumbai 400 085, India*

Chanchal Sow, and P S Anil Kumar

*Department of Physics, Indian Institute of Science, Bangalore 560 012, India*

S Banerjee

*Bhabha Atomic Research Centre, Mumbai 400 085, India*

**Abstract:**

Effects of Fe-substitution on the crystal structure and magnetic correlations of the geometrically frustrated antiferromagnets YBaCo$_{4-x}$Fe$_x$O$_{7+\delta}$ ($x$ = 0, 0.2, 0.4, 0.5, 0.6, and 0.8) have been studied by neutron diffraction, Mössbauer spectroscopy, and ac susceptibility. The compounds YBaCo$_{4-x}$Fe$_x$O$_{7+\delta}$ have special layered crystal structure with alternating Kagomé (6c site) and triangular (2a site) layers along the $c$ axis. Fe$^{3+}$ ions are found to be substituted at both the crystallographic 2$a$ and 6$c$ sites of Co ions. Mössbauer results show a high spin state of Fe$^{3+}$ ions in a tetrahedral coordinate. A reduction in the distortion of the Kagomé lattice has been observed with the Fe-substitution. The correlation length of the short-range antiferromagnetic ordering decreases with the Fe-substitution. The sharpness of the magnetic transition also decreases with the Fe-substitution.

\* Author to whom any correspondence should be addressed.

†Present Address: Helmholtz-Zentrum Berlin für Materialien und Energie, D-14109 Berlin, Germany.

1. **Introduction**

Geometrically frustrated magnetic systems are very interesting both theoretically and experimentally because of their unconventional magnetic ground states. The presence of spin frustrations often leads to a suppression of long-range magnetic ordering, and promotes short-range magnetic correlations due to fluctuations between nearly or totally degenerated magnetic ground states [1]. Strong spin frustrations are found in many complex rare-earth and transition-metal oxides with triangular, Kagomé, spinel, pyrochlore and square crystal lattices. The geometry of the lattice structures causes spin frustrations allowing to tune magnetic properties by controlling the lattice geometry. In this regards, the new class of geometrically frustrated layered mixed-valence cobaltate $YBaCo_4O_{7+\delta}$ and its derivatives with an extended Kagomé structure, belonging to the *swedenborgite* compound family [2-16], are of interest to us.

The compound $YBaCo_4O_{7+\delta}$ have been considered as a model geometrical frustrated system that originates from a special layered crystal structure with an alternating stacking of $CoO_4$ tetrahedral layers of Kagomé ($6c$ site) and triangular ($2a$ site) lattices along the crystallographic $c$-direction (figure 1, and Ref. [17]). The unique crystal structure together with mixed-valence cobalt ions ($Co^{2+}$ and $Co^{3+}$) lead to variety of interesting structural and physical properties. The compound $YBaCo_4O_{7+\delta}$ crystallizes in the trigonal structure (space group $P31c$) [13]. With decreasing temperature, a crystal structural transition occurs from the trigonal phase ($P31c$) to orthorhombic phase ($Pbn2_1$) at ~ 310 K and gives rise to a regime of short-range magnetic order [5, 18]. With further cooling, a magnetic transition to long-range antiferromagnetic (AFM) order state occurs at ~ 110 K [5]. The crystal structural transition lifts geometrical frustrations and facilitates the long-range magnetic ordering. Even in the frustrated trigonal phase (space group $P31c$) of YBaCo$_4$O$_{7+\delta}$, the magnetic Kagomé lattice is distorted and formed with two different sizes of corner sharing equilateral triangles [figure 1(b)]. This indicates that the magnetic correlation in YBaCo$_4$O$_{7+\delta}$ can be tuned by controlling the Kagomé lattice distortions. The compound YBaCo$_4$O$_{7+\delta}$ is found to be highly flexible to the cation substitution allowing to tune its physical properties [13].

The present study focuses in the control of structural distortion with suitable ionic substitution and its role on the magnetic correlations. With this aim, we have substituted Fe for Co and investigated the effect on structural distortion and magnetic properties of YBaCo$_{4-x}$Fe$_x$O$_7$ ($x$ = 0, 0.2, 0.4, 0.5, 0.6, and 0.8) compounds by employing neutron diffraction, Mössbauer spectroscopy, and ac-susceptibility. Neutron diffraction at 22 K reveals a short-range AFM ordering for all compounds with a decrease of correlation length with the Fe-substitution.

2. **Experimental details**

Polycrystalline samples of YBaCo$_{4-x}$Fe$_x$O$_7$ ($x$ = 0, 0.2, 0.4, 0.5, 0.6, and 0.8) were synthesized by a solid state reaction method. Stoichiometric amounts of high purity ($\geq$ 99.99%) Y$_2$O$_3$, BaCO$_3$ Co$_3$O$_4$, and Fe$_3$O$_4$ were taken as the precursors for the reaction. The mixtures of the precursors were initially heated at 1000 ºC for 20 hours in the powder form and then at 1200 ºC in the pellet form for total 60 hours with intermediate grindings. All the heating processes were done in air. The phase purity of the samples was ensured by a room temperature powder x-ray diffraction study, using a rotating anode type Rigaku diffractometer, with a Cu $K_\alpha$ radiation.

Neutron powder diffraction measurements were carried out down to 22 K by using the powder diffractometer II ($\lambda$ =1.249 Å) at Dhruva research reactor, Trombay, India. Cylindrical vanadium sample containers were used for the measurements. A closed cycle helium refrigerator was used to achieve low temperatures. Besides, high resolution neutron diffraction measurements were carried out at 300 K by using the D2B diffractometer ($\lambda$ = 1.590 Å) of the Institut Laue-Langevin (ILL), Grenoble, France, on two representative compounds with $x$ = 0.2 and $x$ = 0.8. The neutron diffraction data were analyzed by the Rietveld method using the FULLPROF program[19].

Mössbauer spectra of $YBaCo_{4-x}Fe_xO_7$ ($x$ = 0.2, 0.4, 0.5, 0.6, and 0.8) samples were recorded at room temperature using a conventional Mössbauer spectrometer (Nucleonix Systems Pvt. Ltd., Hyderabad, India) operated in constant acceleration mode (triangular wave) in transmission geometry. A $^{57}Co$ in Rh matrix of strength 50 mCi γ-ray source was used. An $\alpha$-$^{57}Fe$ metal foil was used at room temperature to calibrate the Doppler velocity and also as the standard sample for the determination of isomer shift ($\delta$). The line width of the calibration spectrum is 0.23 mm s$^{-1}$. Mössbauer spectra were fitted by a least square fit program assuming Lorentzian line shapes. The results of isomer shift are relative to the α-Fe metal foil.

The ac susceptibility ($\chi_{ac}$) measurements were performed by using a commercial CryoBIND susceptometer over the temperature range of 4.2-320 K and in a frequency range of 15-999 Hz.

### 3. Results and Discussion

#### 3.1. Crystal structure at 300 K:

The Rietveld refined neutron diffraction patterns for all compounds ($x$ = 0, 0.2, 0.4, 0.5, 0.6, and 0.8), recorded at 300 K at Trombay, are shown in figures 2 (a-f). The high resolution neutron diffraction patterns, carried out at ILL, are shown in figures 3(a) and 3(b). The Rietveld refinement shows that all compounds crystallize in the trigonal symmetry with the space group $P31c$. The present results are in agreement with the previous reports [15, 16]. With the substitution of Fe, the space group symmetry does not change, however, an increase in the lattice constants, from $a$ = 6.2718(5) Å (for the $x$ = 0) to $a$ =

6.2943(5) Å (for the $x = 0.8$), and from $c = 10.2056(9)$ Å (for the $x = 0$) to $c = 10.2453(8)$ Å (for the $x = 0.8$) has been observed [figure 3(c)]. This leads to an increase in the unit cell volume from $V = 347.66(5)$ Å$^3$ for the $x = 0$ compound to $V = 351.53(5)$ Å$^3$ for the $x = 0.8$ compound [figure 3(c)]. The expansion of the unit cell is due to a higher ionic radius of Fe$^{3+}$ ion (0.645 Å) as compared to that for the Co$^{2+}$ and Co$^{3+}$ ions (0.58 Å and 0.61 Å, respectively) at the tetrahedral coordination [20]. The refined values of lattice constants ($a$ and $c$), unit cell volume, fractional atomic coordinates, isotropic thermal parameters, and occupation numbers of all crystallographic sites are given in table 1.

The variations of the intensity of the (100) and (110) nuclear Bragg peaks with the increasing Fe concentration are shown in figure 2(g). A continuous/monotonous increase of the intensities confirms the substitution of Fe ions at the Co sites. From the Rietveld refinement, it was found that the Fe ions are substituted at both the 6$c$ and 2$a$ sites. A refinement, assuming that the 6$c$ site is fully occupied by Co atoms, and the corresponding 2$a$ site with both the Co and Fe atoms, resulted in a negative thermal displacement ($B_{iso}$) for the 2$a$ site. Hence, a co-occupation of Co and Fe atoms at both the 2$a$ and 6$c$ sites was necessary to obtain reasonable $B_{iso}$ values. A best agreement between observed and calculated patterns was obtained with this configuration. Similar site disorder was also reported for Zn$^{2+}$, Al$^{3+}$, and Ga$^{3+}$ substitutions at the Co-site in YBaCo$_4$O$_7$ [3, 4, 21-24]. The large difference in the neutron coherent scattering lengths for the Co (0.249 × 10$^{-12}$ cm) and Fe (0.945 × 10$^{-12}$ cm) allowed us to determine the occupations of the Co and Fe ions at both the 6$c$ and 2$a$ sites.

The relative site occupancies of the Fe ions (at the 2$a$ and 6$c$ sites) depend on the concentration ($x$), as shown in figure 3(d). At a lower value of $x$, a preference for Fe-substitution at the 2$a$ site has been found. For example, ~79% of Fe ions occupy at the 2$a$ site, and the rest of the Fe ions (~ 21%) are located at the 6$c$ site for the $x = 0.2$ compound. Whereas, for the $x = 0.8$ compound, the relative occupations of Fe ions at the 2$a$ and 6$c$ sites are ~ 46.6% and ~ 53.4%, respectively. These results are consistent with the derived Fe-occupation values from our Mössbauer spectroscopy study (discussed later). Mössbauer spectroscopy study also confirms a three plus (3+) oxidation state for the Fe ions. The above results suggest that $Co^{2+}$ and $Co^{3+}$ ions are located at both the 2*a* and 6*c* sites. In order to understand further, a bond-valence sum (BVS) analysis has been performed by the Fullprof suite program using the output of the refinement of the high resolution neutron diffraction patterns, collected at ILL. The results of the BVS calculations are given in table 2. The BVS analysis suggests that $Co^{2+}$ and $Co^{3+}$ ions are located at both the 2*a* and 6*c* sites. However, there is a site preference of the $Co^{3+}$ ions at the 2*a* site. The results also show that $Ba^{2+}$ was under bonded which is also reported for the similar compounds $YbBaCo_4O_7$ [25] and $TmBaCo_4O_7$ [26]. The relative occupancies of the $Fe^{3+}$ at the 2*a* and 6*c* sites suggest that $Fe^{3+}$ ions are substituted by replacing the $Co^{3+}$ alone at both the sites which is also evident from the Mössbauer results (discussed later).

Now we discuss the effect of the Fe-substitution on the crystal structural parameters of both Kagomé (6*c* site) and triangular (2*a* site) layers which are responsible for dictating the magnetic properties in these compounds. Figure 4 depicts relevant structural parameters, derives from the Trombay data as well as ILL data, as a function of the Fe-concentration. Within the trigonal crystal structure, there are three oxygen sites, labeled as O1, O2, and O3. In a given Kagomé layer, the Co/FeO$_4$ tetrahedra are connected by sharing their corners via O1 or O2 oxygen ions and the magnetic superexchange interactions occur through the (Co/Fe)-O1-(Co/Fe) and (Co/Fe)-O2-(Co/Fe) pathways. The Kagomé lattice is distorted and formed with two different sizes of corner sharing equilateral triangles, where the Co/Fe ions are located at the corners of the triangles. The distorted Kagomé lattice implies superexchange interactions of two different strengths within the plane. The side lengths of these two equilateral triangles in the Kagomé plane are found to be 2.953(8) Å and 3.319(10) Å, respectively, for the parent (*x* = 0) compound. Within the smaller triangle [green triangles in figure 1(b)], the Co ions are connected by the O1 ions, and within the larger triangle [blue triangles in figure 1(b)], the Co ions are connected by the O2 ions. With the Fe-substitution, an increase of the size of the smaller triangle, whereas, a decrease of the size of the larger triangle have been observed [figure 4 (a)]. For the highest Fe-substituted $x = 0.8$ compound, the side lengths are found to be 3.063(9) Å and 3.230(9) Å, respectively. The distortion of the Kagomé lattice ($6c$) can be quantified by the difference between the side lengths (Co-Co distances) of two unequal triangles [shown in figure 4(c)]. It is evident that the distortion decreases with the increasing Fe-concentration.

On the other hand, within a given triangular layer ($2a$), the CoO$_4$ tetrahedra are well separated from each other and connected via YO$_6$ or/and BaO$_{12}$ polyhedrons. Due to a long superexchange pathway via O-Y/Ba-O bridge, a weak magnetic superexchange interaction between Co/Fe ions is expected within the triangular layers. Along the $c$ axis, the Co1 and Co2 magnetic ions (from triangular and Kagomé layers, respectively) are connected via O2/O3 oxygen ions (figure 1). Recent reports show that the magnetic exchange interactions along the $c$ axis play an important role on the magnetic ground state [27]. The direct distances between Co1 and Co2 ions along the $c$ axis, depicted in figure 4(b), reveal two different values. The difference between two Co1-Co2 distances remains almost unchanged with the Fe-substitutions [figure 4(c)].

The effect of Fe-substitution on the CoO$_4$ tetrahedra is discussed below. The variations of the Co-O bond lengths for the $2a$ (Co1) site (triangular layer) and $6c$ (Co2) site (Kagomé layer) with the Fe-substitution are shown in figures 4(d) and 4(e), respectively. For simplicity, the labels for the bond-lengths are shown with respect to Co-ions only (i.e., by omitting Fe-ions). For the Co2O$_4$ tetrahedra ($6c$-site; Kagomé layer), the Co2-O1 bond length increases, whereas, the Co2-O2 and Co2-O3 bond lengths decrease with the Fe-substitution. For the Co1O$_4$ tetrahedra ($2a$ site; triangular layer), all four bond lengths (three equivalent Co1-O3 and Co1-O2) remain unchanged within their error bars. The unequal Co-O bond lengths for both $2a$ and $6c$ sites indicate a distortion of the tetrahedron. The distortion of the tetrahedron may be characterized by the quantity $\Delta$ as $\Delta = \left(\frac{1}{4}\right)\sum_i^4\{(d_i - <d>)/<d>\}^2$

(1)

where, $d_i$ is the $i^{th}$ bond length and $<d>$ is the average of the bond lengths. The variation of distortion of the tetrahedron as a function of Fe concentration is shown in figure 4(f). For both the 2a and 6c sites, the distortion of the tetrahedron decreases with the Fe substitution. However, the variation is significantly larger for the Kagomé layer.

### 3.2. Mössbauer study:

In order to confirm the Fe occupations at both the 2a and 6c sites, as well as oxidation and spin states of the Fe ions in the Fe-substituted compounds YBaCo$_{4-x}$Fe$_x$O$_7$ ($x$ = 0.2, 0.4, 0.5, 0.6, and 0.8), Mössbauer spectra (figure 5) were recorded at room temperature. The observed spectra for all compounds could be fitted with two doublets corresponding to the two crystallographic sites (Co1/2a and Co2/6c sites). This reconfirms the presence of Fe ions at both the 2a and 6c sites, as found in the neutron diffraction study. The earlier report for the YBaCo$_{3.96}$Fe$_{0.04}$O$_{7.02}$ compound with a small Fe concentration also showed the presence of two doublets at room temperature [8]. For the present compounds, the values of the isomer shift (IS), quadruple splitting ($\Delta E_Q$), relative intensities ($R_A$), and line-width ($\Gamma$) for both components are given in table 3.

The values of the IS for both sites are very similar (~ 0.19), and are typical for the high-spin Fe$^{3+}$ in a tetrahedral coordination, as reported for other layered oxide systems [28]. The derived IS values are also in good agreement with the reported values (0.18 and 0.19 mm s$^{-1}$ for two sites), by Tsipis *et al.* [8], for the YBaCo$_{3.96}$Fe$_{0.04}$O$_{7.02}$ compound. The variation of IS with Fe-concentration is shown in figure 6(a). Our Mössbauer data do not show any evidence of Fe$^{2+}$ oxidation state in the Fe-substituted compounds as the IS values for the Fe$^{2+}$ ions at tetrahedral and octahedral coordinates are expected to be ≥ 0.8 mm/s and ≥ 1.0 mm/s, respectively [29].

The $\Delta E_Q$ values are found to be different for the 2a and 6c sites. Since, $\Delta E_Q$ occurs due to an interaction between nuclear quadruple moment and electric field gradient produced by surrounding ions (oxygen ions for the present case), the different values of $\Delta E_Q$ for the two sites suggest different local charge environments for the 2$a$ and 6$c$ sites. The higher value of $\Delta E_Q$ for the 6$c$ site as compared to that for the other site (2$a$ site) indicates the presence of a relatively larger crystal field gradient at the 6$c$ site. For a tetrahedral site, a larger field gradient is expected when a distortion occurs in the tetrahedron. In the present case, the tetrahedra for the 6$c$ site (Kagomé layer) are more distorted than that for the 2$a$ site (triangular layer), as evident from our room temperature neutron diffraction results [figure 4(f)]. The observed decreasing trend of the $\Delta E_Q$ values for both 6$c$ and 2$a$ sites [figure 6(b)] is consistent with the results of neutron diffraction study where less distorted tetrahedra were found with increasing the Fe-concentration [figure 4(f)].

The variations of the relative intensities (absorption), corresponding to the occupations of the Fe ions at the 2$a$ and 6$c$ sites, are shown in figure 6(c). The observed variations agree with the neutron diffraction results [figure 3(d)]. The fractions of the $Fe^{3+}$ ions at the 2$a$ and 6$c$ sites are derived from the relative intensity (absorption) of the two doublets, and given in the last two columns in table 3. The fractions of the $Fe^{3+}$ ions (table 3) indicate that the concentrations of $Fe^{3+}$ increase at both the 2$a$ and 6$c$ sites with Fe-substitution. A preference of Fe-substitution at the 2$a$ site is evident. With substitution of Fe, the concentration of the $Fe^{3+}$ ions at the 2$a$ site saturates at a value ~ 0.34, and then the remaining $Fe^{3+}$ ions occupy at the 6$c$ site. This shows that the $Fe^{3+}$ ions are substituted by replacing $Co^{3+}$ ions alone at both the 2$a$ and 6$c$ sites where the concentration of $Fe^{3+}$ saturates at the 2$a$ site when all the $Co^{3+}$ ions are replaced by $Fe^{3+}$ ions. This also suggests that the concentration of the $Co^{3+}$ ions at the 2$a$ site is ~ 0.34.

### 3.3 ac susceptibility:

In order to understand the nature of magnetic ordering in the Fe-substituted compounds, we have carried out ac-susceptibility measurements. The temperature dependent real part of the ac-susceptibility curves [$\chi(T)$] for all compounds are shown in figure 7(a). For the parent compound YBaCo$_4$O$_7$, with lowering of temperature, $\chi(T)$ curve shows a peak at a temperature ($T_P$) ~ 70 K, followed by a broad hump over the temperature range of 15–60 K. A similar type of susceptibility behaviour for the parent compound was reported for both polycrystalline [2, 17] and single crystal (*ab* plane susceptibility) [6] samples. With Fe-substitution, the peak temperature is found to decrease. For the highest Fe-substituted compound ($x = 0.8$), the $T_P$ becomes ~ 59 K. The variation of the $T_P$ with the Fe-concentration is shown in the inset of figure 7(a). Here, a relatively sharp decrease in $T_P$ over the lower concentration range ($x < 0.4$) is evident. With increasing Fe-concentration, a broadening of the $\chi(T)$ peak at $T_P$ has also been observed, suggesting a decrease in the sharpness of the magnetic transition. It is also observed that the height of the broad peak reduces significantly with Fe-substitution. To understand the nature of the magnetic ground state further, we have performed a frequency dependent ac-susceptibility study on the highest Fe-substituted compound ($x = 0.8$). The $\chi(T)$ curves at different frequencies (over 15–999 Hz) are shown in figure 7(b). An enlarged view of the $\chi(T)$ curves around the peak temperature ($T_P$) is shown in the inset of figure 7(b). No frequency dependent peak shifting has been found, confirming the presence of an AFM correlation in the $x = 0.8$ compound, as found in the parent compound [17].

**3.4. Magnetic correlations at 22 K:**

Figure 8(a) shows the observed neutron diffraction pattern (solid circles) recorded at 22 K for the $x = 0.2$ compound, and a calculated pattern (solid curve) by considering only the nuclear phase (trigonal crystal structure with space group *P*31*c*). An additional broad (not the instrumental resolution limited) asymmetric type peak at $Q$ ~ 1.35 Å$^{-1}$ has been found, suggesting the presence of an AFM spin-spin correlation at this temperature. It also confirms the absence of a true long-range magnetic ordering. With Fe-substitution, the evolution of the magnetic peak is shown in figure 8 (b). For all compounds, the same broad peak is present at $Q$ ~ 1.35 Å$^{-1}$ at 22 K which signifies a similar type of spin arrangements and periodicity. Nevertheless, with the increasing Fe–concentration, the peak becomes broader [figure 8(c-f)] indicating a decrease in the spin-spin correlation length. The values of the peak widths, derived from the fit of the Lorentzian functions, are 0.0691 Å$^{-1}$ for $x = 0$ and 0.162 Å$^{-1}$ for $x = 0.8$, which correspond to the spin-spin correlation lengths of 14.5 Å and 6.2 Å, respectively. The intensity of the magnetic peak also decreases with the increasing Fe-concentration. The magnetic peaks start to appear below ~ 130 K in the neutron diffraction patterns for all compounds (not shown here) [17].

Now we discuss below the possible magnetic correlations in these compounds. Similar asymmetric diffuse neutron scattering, indication of a broken long-range magnetic ordering, was reported for YBaCo$_4$O$_{7.0}$ above its long-range magnetic ordering temperature $T_N$ (~ 110 K) [5], and was attributed to the distinct pattern of short-range magnetic order derived from a unique magnetic exchange topology of linked trigonal bipyramids of Co [15, 18]. The diffuse neutron scattering was also reported for the isostructural compound Y$_{0.5}$Ca$_{0.5}$BaCo$_4$O$_7$ [30] as well as YBaCo$_4$O$_{7+\delta}$ [16, 17] with $\delta > 0.08$, however, over a wide temperature range below ~ 110 K down to lowest measured temperatures (6 K and 2 K, respectively) without any transition to a long-range magnetic order. For Y$_{0.5}$Ca$_{0.5}$BaCo$_4$O$_7$, the broken long-range magnetic order was attributed to the random distribution of Ca atoms that causes a local structural disorder and interferes with the magnetic exchange pathways [30, 31]. For YBaCo$_4$O$_{7+\delta}$ ($\delta > 0.08$), an excess oxygen content preserves the geometrically frustrated trigonal structure down to 6 K, and reveals only the short-range magnetic order [16]. For all the present compounds, a short-range magnetic ordering is evident from the broad peak in the neutron diffraction patterns down to 22 K. In agreement with the present neutron diffraction results, a broad magnetic peak in the polarized neutron diffuse scattering pattern was reported for the YBaCo$_3$FeO$_7$ compound [32]. The nature of the magnetic ordering was linked to a quasi-one dimensional partially ordered AFM state.

Now, we compare the magnetic ordering in the present Fe-substituted compounds with the earlier reports on the Fe-substituted compounds. A single crystal Mössbauer study on YBaCo$_3$FeO$_7$ revealed two spin freezing temperatures of $T_{c1} = 590$ K and $T_{c2} = 50$ K corresponding to the spin freezing at the Kagomé layers and triangular layers, respectively [32]. However, a Mössbauer study on powder compounds [YBaCo$_{4-x}$Fe$_x$O$_{7+\delta}$ ($x$ = 0–0.8)] reported single spin freezing temperature ~ 70 K [33]. The present neutron diffraction study shows the onset of the magnetic ordering around 130 K for all compounds, however, with no observable change in the magnetic diffraction peak around $T_{c2}$. The present ac susceptibility data (figure 7) do show a peak around $T_{c2}$. Moreover, at 22 K, the monotonous decrease of the spin-spin correlation length with the Fe-substitution reveals the increase of spin-frustration. In agreement with our observation, the increase of spin-frustration is also evident from the increase of the Curie-Weiss temperature ($\theta_{C-W}$ = -907, -1150, -1304, and -2000 K for $x$ =0, 0.04, 0.8, and 1.0, respectively) with the Fe-substitution [32, 33]. The increase of frustration may be correlated to the decrease of the distortion in the Kagomé plane (decrease in the difference between the Co2-Co2 distances within the $ab$ plane) with the Fe-substitution [figure 4(c)]. However, one cannot rule out the contribution from a possible spin disorder effect due to the Fe-substitution.

4. **Summary and Conclusion**

In summary, the effects of Fe-substitution on the crystal structural and magnetic properties of the geometrical frustrated mixed valence (Co$^{2+}$ and Co$^{3+}$) cobaltates YBaCo$_{4-x}$Fe$_x$O$_{7+\delta}$ ($x$ = 0, 0.2, 0.4, 0.5, 0.6, and 0.8), consisting of an alternating the Kagomé and triangular layers, are reported. Neutron diffraction and Mössbauer spectroscopy results show that the Fe ions are substituted at both Kagomé (6$c$ site) and triangular (2$a$ site) layers. All Fe ions are found to be in the three plus (Fe$^{3+}$) oxidation state and substituted at the place of Co$^{3+}$ ions alone. With the Fe-substitution, the crystal symmetry (trigonal symmetry with the space group $P31c$) remained unchanged; however, an expansion of the unit cell volume occurs due to higher ionic radius of the Fe$^{3+}$ than that of the both Co$^{2+}$ and Co$^{3+}$ ions at the tetrahedral coordination.. With the Fe-substitution, a decrease of the Kagomé lattice distortion occurs that leads to a decrease of short-range spin-spin correlation length. The decrease of spin-spin correlation length indicates the increase of spin frustrations with the Fe-substitution which is in agreement with the previous reports on the increase of Curie-Weiss temperature.

**Acknowledgment**

AKB acknowledges the help rendered by A. B. Shinde in performing neutron diffraction experiments. AKB and SMY acknowledge the facility used for the high resolution neutron diffraction measurements at Instititute Laue Langevin, Grenoble, France, and also thank C. Ritter for the measurements.

**References**

[1]  Moessner R and Ramirez A P 2006 *Phys. Today* **59** 24

[2]  Valldor M and Andersson M 2002 *Solid State Sci.* **4** 923

[3]  Valldor M 2004 *Solid State Sci.* **6** 251

[4]  Valldor M, Hollmann N, Hemberger J and Mydosh J A 2008 *Phys. Rev. B* **78** 024408

[5]  Chapon L C, Radaelli P G, Zheng H and Mitchell J F 2006 *Phys. Rev. B* **74** 172401

[6]  Soda M, Yasui Y, Moyushi T, Sato M, Igawa N and Kakurai K 2006 *J. Phys. Soc. Japan* **75** 054707

[7]  Tsipis E V, Khalyavin D D, Shiryaev S V, Redkina K S and Nunez P 2005 *Mater. Chem. Phys.* **92** 33

[8]  Tsipis E V, Waerenborgh J C, Avdeev M and Kharton V V 2009 *J. Solid State Chem.* **182** 640

[9]  Hao H, Chen C, Pan L, Gao J and Hu X 2007 *Physica B* **387** 98

[10] Gatalskaya V I, Dabkowska H, Dube P, Greedan J E and Shiryaev S V 2007 *Phys. Solid State* **49** 1125

[11] Maignan A, Caignaert V, Pelloquin D, Hébert S, Pralong V, Hejtmanek J and Khomskii D 2006 *Phys. Rev. B* **74** 165110

[12] Valldor M 2007 *New Topics in Condensed Matter Research,* ed L R Velle (New York: Nova Science Publiser) p 75

[13] Parkkima O and Karppinen M 2014 *Eur. J. Inorg. Chem.* 4056

[14] Hao H, He Q, Cheng Y and Zhao L 2014 *J. Phys. Chem. Solid* **75** 495

[15] Manuel P, Chapon L C, Radaelli P G, Zheng H and Mitchell J F 2009 *Phys. Rev. Lett.* **103** 037202

[16] Avci S, Chmaissem O, Zheng H, Huq A, Manuel P and Mitchell J F 2013 *Chem. Mater.* **25** 4188

[17] Bera A K, Yusuf S M and Banerjee S 2012 *Solid State Sc.* **16** 57

[18] Khalyavin D D, Manuel P, Ouladdiaf B, Huq A, Stephens P W, Zheng H, Mitchell J F and Chapon L C 2011 *Phys. Rev. B* **83** 094412

[19] http://www.ill.eu/sites/fullprof/.

[20] Shannon R D 1976 *Acta Cryst.* **A32** 751

[21] Valldor M 2004 *J. Phys.: Condens. Matter* **16** 9209

[22] Hao H, Zhao L, Hu X and Hou H 2009 *J.Thermal Analysis and Calorimetry* **95** 585

[23] Hao H, Zhang X, He Q, Chen C and Hu X 2007 *Solid State Commun.* **141** 591

[24] Valldor M 2005 *Solid State Sci.* **7** 1163

[25] Huq A, Mitchell J F, Zheng H, Chapon L C, Radaelli P G, Knight K S and Stephens P W 2006 *J. Solid State Chem.* **179** 1136

[26] Khalyavin D D, Chapon L C, Radaelli P G, Zheng H and Mitchell J F 2009 *Phys. Rev. B* **80** 144107

[27] Khalyavin D D, Manuel P, Mitchell J F and Chapon L C 2010 *Phys. Rev. B* **82** 094401

[28] Waerenborgh J C, Rojas D P, Vyshatko N P, Shaula A L, Kharton V V, Marozau I P and Naumovich E N 2003 *Mater. Letters* **57** 4388

[29] Greenwood N N and Gibb T C 1971 *Mössouber Spectroscopy* (London: Chapman and Hall Ltd.)

[30] Schweika W, Valldor M and Lemmens P 2007 *Phys. Rev. Lett.* **98** 067201

[31] Stewart J R, Ehlers G, Mutka H, Fouquet P, Payen C and Lortz R 2011 *Phys. Rev. B* **83** 024405

[32] Valldor M, Hermann R P, Wuttke J, Zamponi M and Schweika W 2011 *Phys. Rev. B* **84** 224426

[33] Waerenborgh J C, Tsipis E V, Pereira L C J, Avdeev M, Naumovich E N and Kharton V V 2012 *Dalton Trans.* **41** 667

**Table 1.** The Rietveld refined lattice constants (*a* and *c*), unit cell volumes (*V*), fractional atomic coordinates, isotropic thermal parameters ($B_{iso}$), site occupancies (Occ.), $\chi^2$ and other agreement factors such as, profile factor ($R_p$), weighted profile factor ($R_{wp}$), expected weighted profile factor ($R_{exp}$) for the compounds $YBaCo_{4-x}Fe_xO_7$ ($x$ = 0, 0.2, 0.4, 0.5, 0.6, and 0.8) at 300 K.

| | $x = 0$ | $x = 0.2$ | $x = 0.4$ | $x = 0.5$ | $x = 0.6$ | $x = 0.8$ |
|---|---|---|---|---|---|---|
| *a* (Å) | 6.2718(5) | 6.2834(8) | 6.2877(6) | 6.2872(4) | 6.2907(4) | 6.2943(5) |
| *c* (Å) | 10.2056(9) | 10.2297(4) | 10.2350(1) | 10.2288(8) | 10.2375(8) | 10.2453(8) |
| *V* (Å$^3$) | 347.66(5) | 349.77(8) | 350.43(6) | 350.16(4) | 350.86(4) | 351.53(5) |
| **Y** | | | | | | |
| 2*b* (2/3, 1/3, z) | | | | | | |
| z/c | 0.8549(6) | 0.8670(2) | 0.8477 (4) | 0.8463(4) | 0.8454 (3) | 0.8437 (2) |
| $B_{iso}$ (Å$^2$) | 1.0(2) | 0.8(3) | 0.9(2) | 0.9(1) | 1.0(2) | 1.0(3) |
| Occ. | 1.00(1) | 1.01(1) | 1.00(1) | 1.00(1) | 1.01(1) | 1.0(1) |
| **Ba** | | | | | | |
| 2*b* (2/3, 1/3, z) | | | | | | |
| z/c | 0.4763(7) | 0.4878(4) | 0.4670 (1) | 0.4682(3) | 0.4682 (1) | 0.4680(3) |
| $B_{iso}$ (Å$^2$) | 1.3(1) | 1.5(2) | 2.0(2) | 2.0(1) | 1.4(1) | 1.5(2) |
| Occ. | 1.00(4) | 1.1(5) | 1.01(2) | 1.02(2) | 1.01(3) | 1.01(3) |
| **Co1/Fe1** | | | | | | |
| 2*a* (0, 0, z) | | | | | | |
| z/c | 0.4285 (2) | 0.4326(8) | 0.4099 (4) | 0.4131(4) | 0.4161 (3) | 0.4090 (4) |
| $B_{iso}$ (Å$^2$) | 1.1(4) | 1.1(4) | 0.9(3)/ 0.9(1) | 1.2(1)/ 1.2(1) | 1.0(2)/1.0(2) | 1.1(1)/ 1.1(1) |
| Occ. | 1.00(2) | 0.84(2)/ 0.16(2) | 0.73(1)/ 0.27(1) | 0.73(1)/ 0.27(1) | 0.67(1)/ 0.33(1) | 0.63(1)/ 0.37(1) |
| **Co2/Fe2** | | | | | | |
| 6*c* (x, y, z) | | | | | | |
| x/a | 0.1698(3) | 0.1742(2) | 0.1708 (9) | 0.1689(9) | 0.1681(3) | 0.1640(2) |
| y/b | 0.8175(4) | 0.8220(5) | 0.8255(5) | 0.8226(4) | 0.8235(3) | 0.8229(5) |
| z/c | 0.6724(8) | 0.6859 (5) | 0.6622(8) | 0.6616(6) | 0.6598(6) | 0.6592(3) |
| $B_{iso}$ (Å$^2$) | 0.5(1) | 0.4(1)/0.4(1) | 0.5(1)/ 0.5(1) | 0.5(1)/ 0.5(3) | 0.8(1)/ 0.8(1) | 0.7(1)/0.7(1) |
| Occ. | 0.99(1) | 0.98(1)/ 0.02(1) | 0.96(1)/ 0.04(1) | 0.92(1)/ 0.08(1) | 0.91(1)/ 0.09(1) | 0.86(1)/ 0.14(1) |
| **O1** | | | | | | |
| 6*c* (x, y, z) | | | | | | |
| x/a | 0.5095 (2) | 0.5115(4) | 0.5078 (7) | 0.5065(5) | 0.5079 (7) | 0.5122(8) |
| y/b | 0.5016(2) | 0.5014(6) | 0.4972(3) | 0.4989(4) | 0.4985(9) | 0.5081(3) |
| z/c | 0.7353(5) | 0.7431 (7) | 0.7214(6) | 0.7249(8) | 0.7234(4) | 0.7191(6) |
| $B_{iso}$ (Å$^2$) | 2.8(8) | 2.7(3) | 2.6(1) | 2.5(7) | 2.5(8) | 2.8(1) |
| Occ. | 1.00(1) | 1.00(3) | 1.00(1) | 0.98(1) | 1.02(3) | 1.04(1) |
| **O2** | | | | | | |
| 2a (0, 0, z) | | | | | | |
| z/c | 0.2308(6) | 0.2512(8) | 0.2250(2) | 0.2268(8) | 0.2287(7) | 0.2207(3) |
| $B_{iso}$ (Å$^2$) | 1.1(1) | 1.4(1) | 1.5(1) | 1.1(2) | 1.1(3) | 1.4(1) |
| Occ. | 1.01(1) | 1.09(4) | 1.02(2) | 1.03(2) | 1.03(2) | 1.02(2) |
| **O3** | | | | | | |
| 6c (x, y, z) | | | | | | |
| x/a | 0.1433(6) | 0.1441(4) | 0.1456(4) | 0.1462(5)) | 0.1387(7) | 0.1432(8) |
| y/b | 0.8132(7) | 0.8218(6) | 0.8254(2) | 0.8211(4) | 0.8132(2) | 0.8175(2) |
| z/c | 0.4748(6) | 0.4845(8) | 0.4624(9) | 0.4663(8) | 0.4668(7) | 0.4642(3) |
| $B_{iso}$ (Å$^2$) | 1.6(9) | 1.4(1) | 1.9(6) | 1.3(7) | 1.4(2) | 1.7(1) |
| Occ. | 1.01(1) | 1.04(3) | 1.02(1) | 1.03(1) | 1.06(4) | 0.99(1) |
| $\chi^2$ | 2.19% | 2.34% | 1.95% | 2.09% | 2.27% | 2.39% |
| $R_p$ | 3.11% | 3.16% | 2.78% | 2.95% | 3.31% | 3.18% |
| $R_{wp}$ | 3.96% | 4.02% | 3.53% | 3.76% | 4.22% | 4.07% |
| $R_{exp}$ | 2.68% | 2.62% | 2.53% | 2.60% | 2.80% | 2.63% |

**Table 2.** Results of the bond-valence sum (BVS) calculation for the cations from the high resolution diffraction data at 300 K.

| Atoms | Site | BVS | |
|---|---|---|---|
| | | x = 0.2 | x = 0.8 |
| Y | 2b | 3.47(4) | 3.42(3) |
| Ba | 2b | 1.30(1) | 1.24(1) |
| Co1 | 2a | 2.75(4) | 2.45(1) |
| Fe1 | 2a | 3.29(4) | 2.99(4) |
| Co2 | 6c | 2.15(5) | 2.114) |
| Fe2 | 6c | 2.58(6) | 2.53(3) |

**Table 3.** The Mössbauer parameters for the compounds $YBaCo_{4-x}Fe_xO_7$ ($x$ = 0.2, 0.4, 0.5, 0.6, and 0.8) at room temperature.

| Fe ($x$) | IS1 (mm s$^{-1}$) | $\Delta E_Q$1 (mm s$^{-1}$) | IS2 (mm s$^{-1}$) | $\Delta E_Q$2 (mm s$^{-1}$) | $R_A$1 (%) | $R_A$2 (%) | $\Gamma$1 (mm s$^{-1}$) | $\Gamma$2 (mm s$^{-1}$) | 2a | 6c |
|---|---|---|---|---|---|---|---|---|---|---|
| 0.2 | 0.193(4) | 0.423(10) | 0.191(5) | 0.764(10) | 91.36(3) | 8.64(3) | 0.293(7) | 0.233(7) | 0.18 | 0.02 |
| 0.4 | 0.191(4) | 0.417(9) | 0.200(4) | 0.723(8) | 80.11(2) | 19.89(2) | 0.30 | 0.34 | 0.32 | 0.08 |
| 0.5 | 0.192(4) | 0.398(9) | 0.195(5) | 0.739(8) | 68.78(2) | 31.22(2) | 0.28 | 0.33 | 0.34 | 0.16 |
| 0.6 | 0.190(5) | 0.350(8) | 0.196(5) | 0.614(9) | 55.79(1) | 44.21(1) | 0.24 | 0.33 | 0.33 | 0.27 |
| 0.8 | 0.189(4) | 0.376(9) | 0.195(4) | 0.676(9) | 42.72(2) | 57.28(2) | 0.25 | 0.38 | 0.34 | 0.46 |

**Figures**

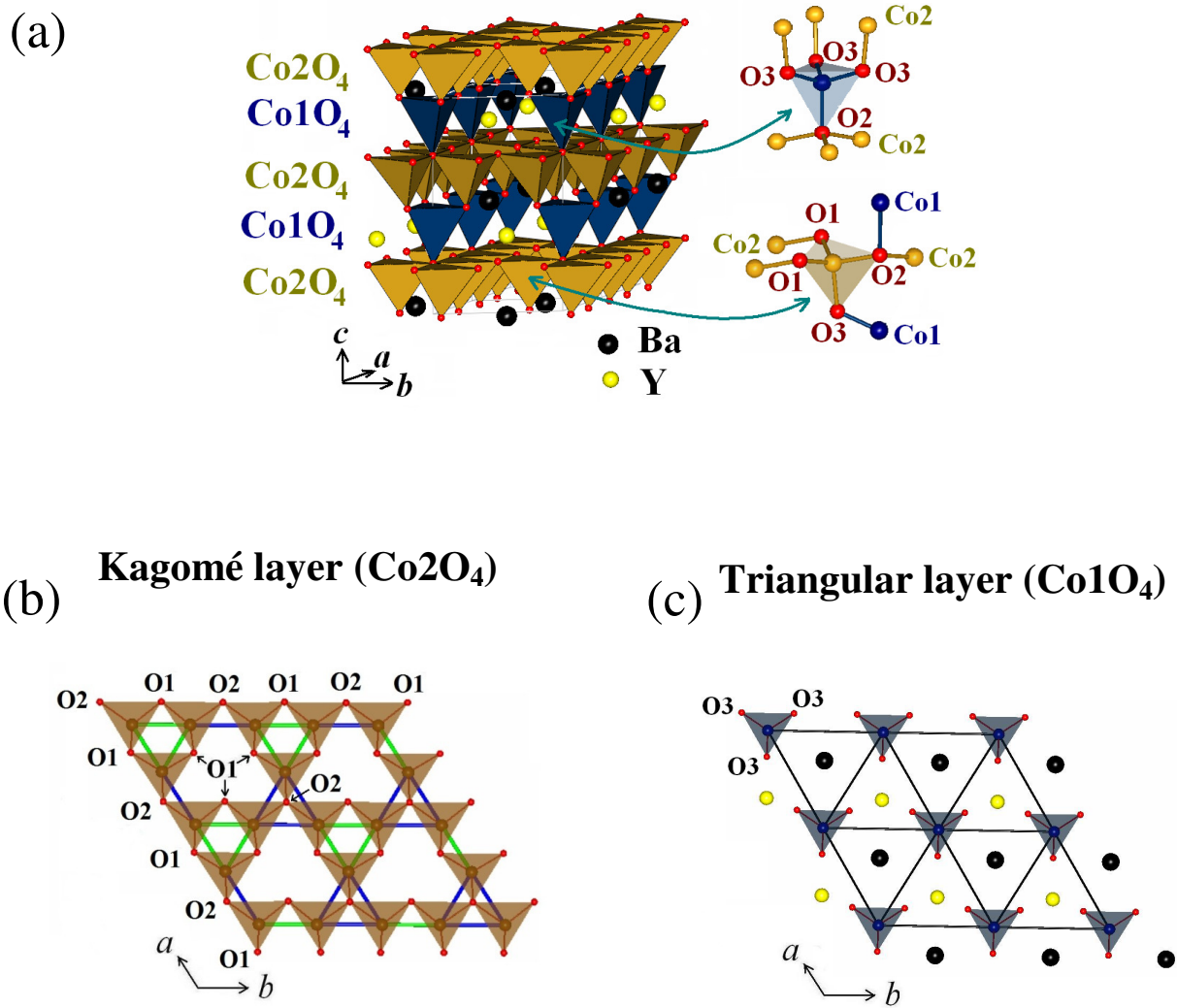

**Figure 1.** (Colour online) (a) The layered type crystal structure of the YBaCo$_{4-x}$Fe$_x$O$_7$ compounds. The local coordinations for both the Kagomé and triangular layers have been shown. The geometrical arrangements of the CoO$_4$ tetrahedra within a given *ab* plane for the (b) Kagomé layers (6*c* site) and (c) triangular layers (2*a* site). Within the Kagomé lattice, two equilateral triangles with different sizes are shown by green and blue lines, respectively. The size of green tringles (connected by O1 ions) is smaller than that of the blue triangles (connected by O2 ions).

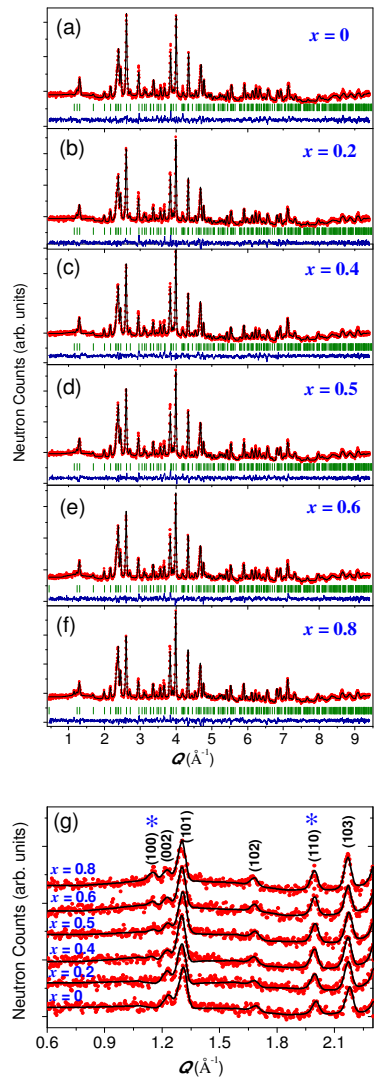

**Figure 2.** (Colour online) (a-f) Observed (filled circles) and calculated (solid lines) neutron diffraction patterns, at 300 K, for YBaCo$_{4-x}$Fe$_x$O$_7$ with $x$ = 0, 0.2, 0.4, 0.5, 0.6, and 0.8. The diffraction patterns were recorded using the powder diffractometer II at Dhruva research reactor, Trombay, INDIA. Solid line at the bottom of each panel shows the difference between observed and calculated patterns. Vertical lines show the position of Bragg peaks. (g) An enlarge view of low $Q$-region of the neutron diffraction patterns for all YBaCo$_{4-x}$Fe$_x$O$_7$ ($x$ = 0, 0.2, 0.4, 0.5, 0.6, and 0.8) compounds, measured at 300 K. The (*hkl*) values for observed peaks are also listed. With the increase of the Fe-concentration, an increase in the intensity of the (100) and (110) nuclear Bragg peaks is observed, and these peaks are marked with asterisks.

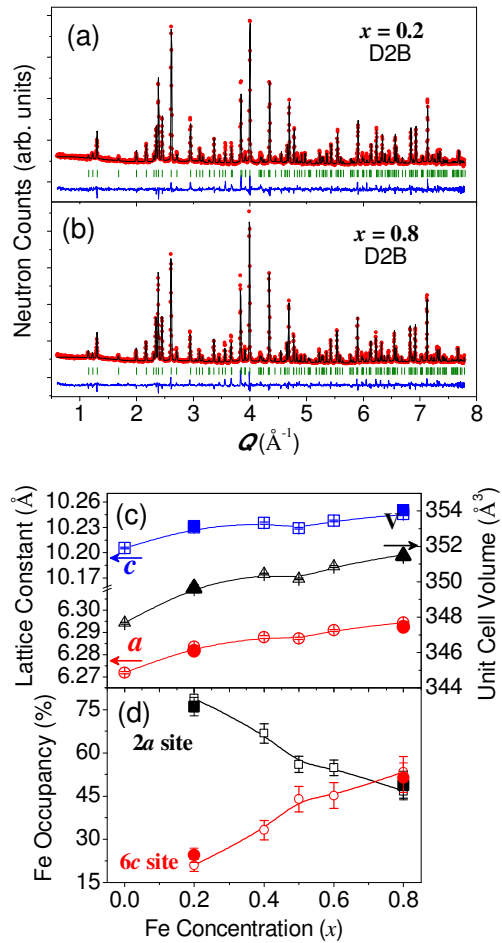

**Figure 3.** (Colour online) The high resolution neutron diffraction patterns, at 300 K, for two representative samples (a) $x = 0.2$ and (b) $x = 0.8$, recorded using the D2B at the ILL, Grenoble, France. Observed and calculated patterns are shown by filled circles and solid lines, respectively. Solid line at the bottom of each panel shows the difference between observed and calculated patterns. Vertical lines show the position of Bragg peaks. (c) The Fe concentration ($x$) dependent lattice constants ($a$ and $c$) and unit cell volume ($V$) for the YBaCo$_{4-x}$Fe$_x$O$_7$ compounds. (d) The relative occupation of the Fe$^{3+}$ ions at the $2a$ and $6c$ sites as a function of Fe concentration ($x$) for the YBaCo$_{4-x}$Fe$_x$O$_7$ compounds. The filled symbols in (c) and (d) show the results from the high resolution neutron diffraction study at the ILL on the $x = 0.2$ and 0.8 compounds.

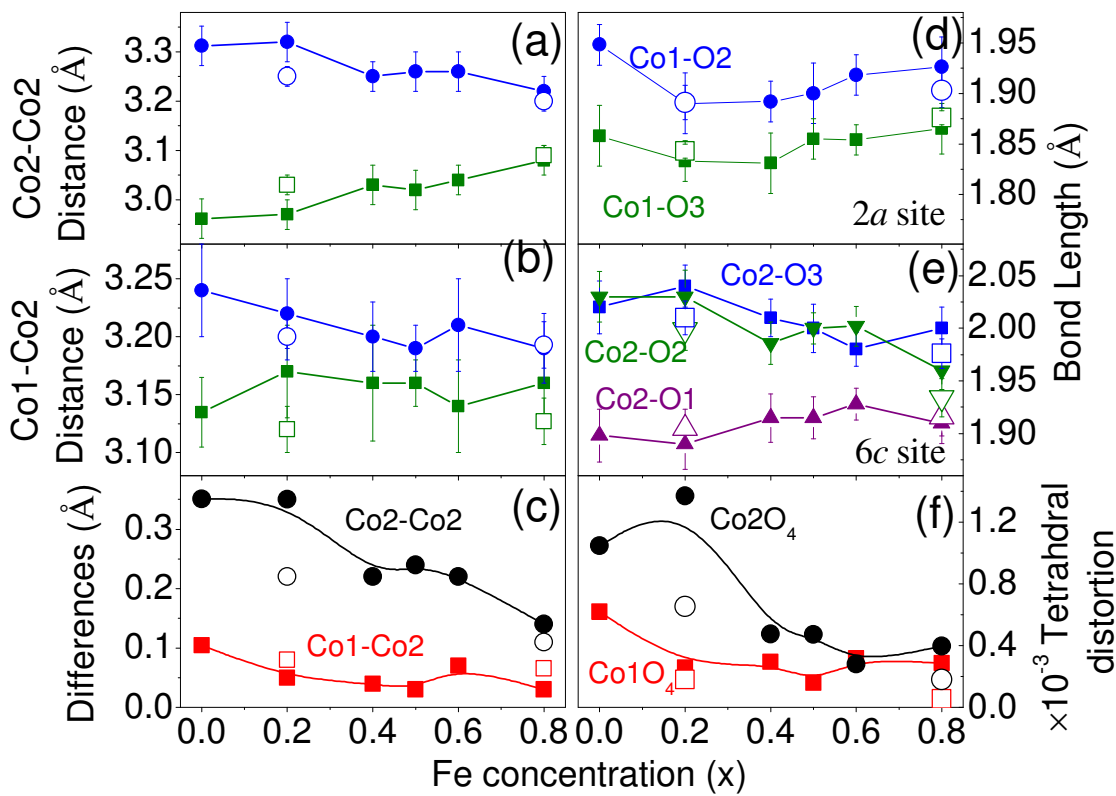

**Figure 4.** (Colour online) The structural parameters as a function of the Fe-concentration for the YBaCo$_{4-x}$Fe$_x$O$_7$ compounds. (a) The Co2-Co2 direct distances for the 6$c$ site. (b) The Co1-Co2 direct distances along the $c$ axis. (c) The differences between two Co2-Co2 and two Co1-Co2 direct distances. (d) and (e) The Co-O bond lengths for the 2$a$ site and the 6$c$ site, respectively, as a function of Fe-concentration. (f) Distortion of the tetrahedron as a function of the Fe-concentration for both the 2$a$ and 6$c$ sites. The results derived from the high resolution neutron diffraction study at the ILL for the $x$ = 0.2 and 0.8 compounds are shown by the open symbols in all panels.

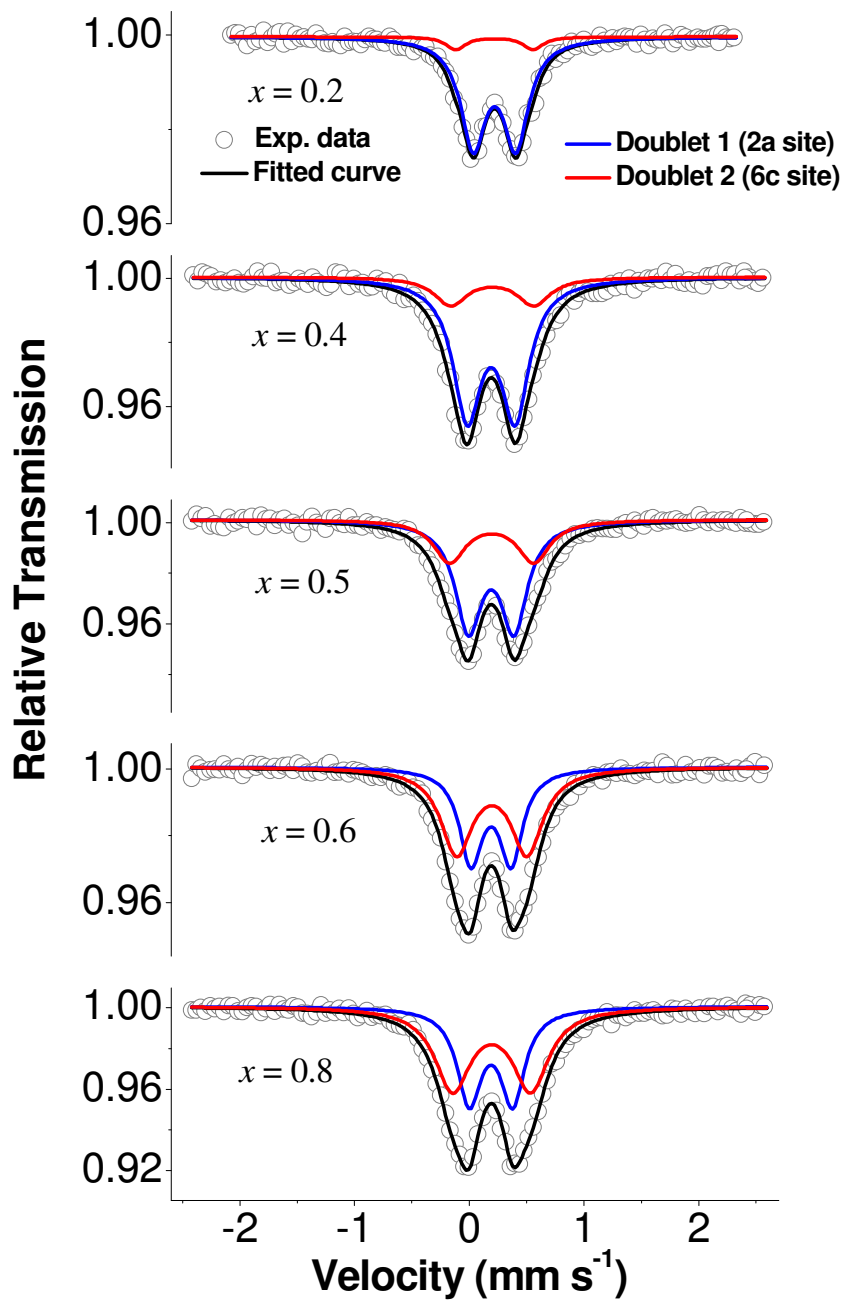

**Figure 5.** (Colour online) The Mössbauer spectra at room temperature for the Fe-substituted YBaCo$_{4-x}$Fe$_x$O$_7$ ($x$ = 0.2, 0.4, 0.5, 0.6, and 0.8) compounds. The solid lines are the least square fitted curves considering two doublets.

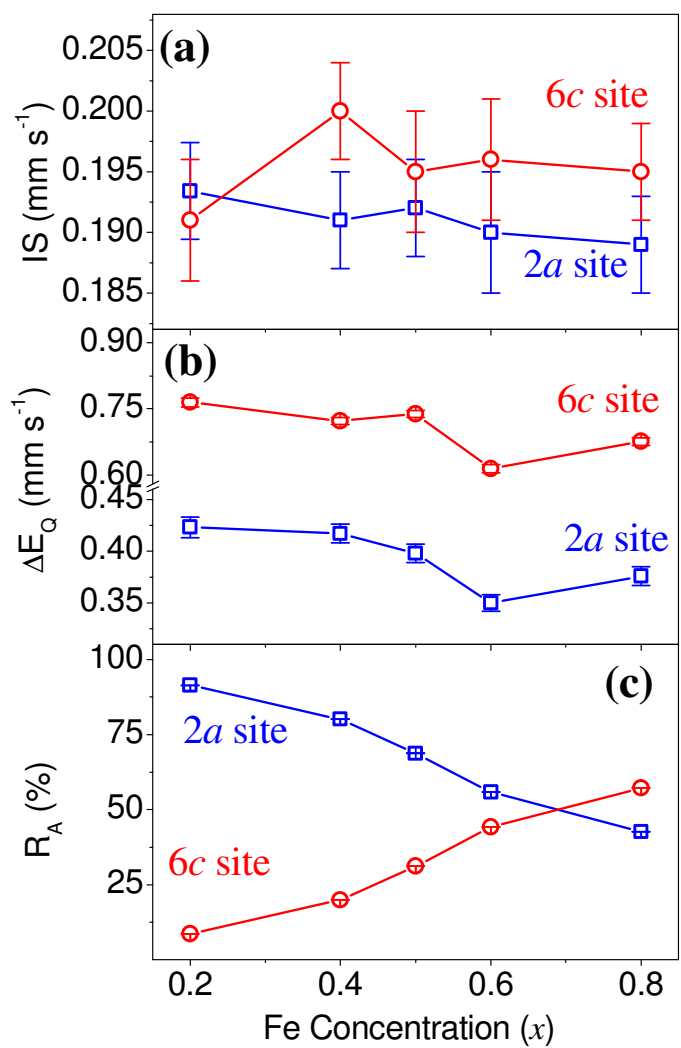

**Figure 6.** (Colour online) The Fe-concentration ($x$) dependent (a) isomer shift (IS), (b) quadruple splitting ($\Delta E_Q$), and (c) Relative intensity ($R_A$) for the 2$a$ and 6$c$ sites derived from the analysis of the Mössbauer spectra at room temperature.

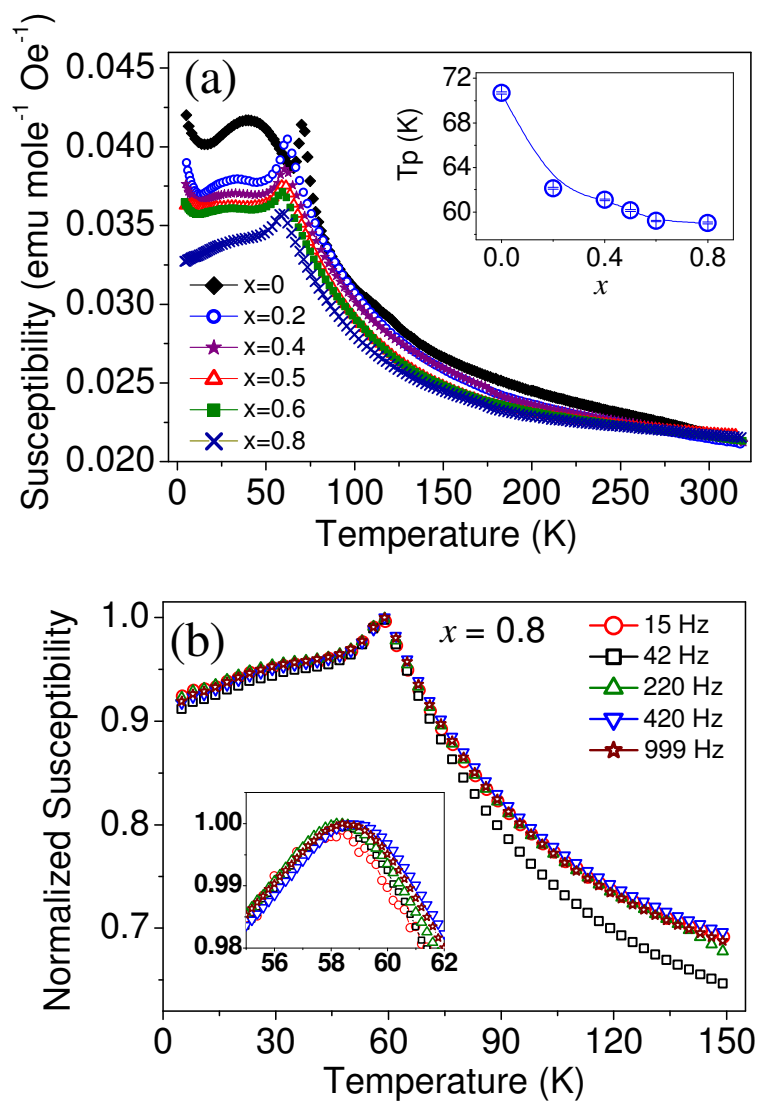

**Figure 7.** (Colour online) (a) The temperature dependent real part of the ac susceptibility curves for the YBaCo$_{4-x}$Fe$_x$O$_7$ ($x$ = 0, 0.2, 0.4, 0.5, 0.6, and 0.8) compounds, measured under 420 Hz frequency. The inset shows the variation of peak temperature ($T_p$) with the Fe-concentration ($x$). (b) The $\chi(T)$ curves for the highest Fe-substituted compound $x$ = 0.8 under 15, 42, 220, 420, and 999 Hz ac frequencies. The inset shows an enlarge view around the peak temperature.

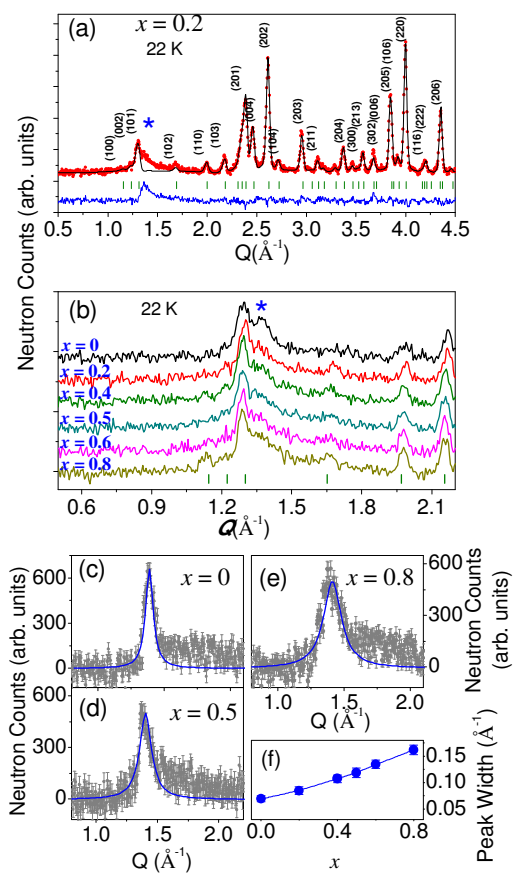

**Figure 8.** (Colour online) (a) Experimentally (circles) observed neutron diffraction pattern at 22 K and a calculated (solid line) curve by considering only nuclear phase (crystal structure) for the $YBaCo_{3.8}Fe_{0.2}O_7$ ($x = 0.2$) compound. Solid line at the bottom shows the difference between observed and calculated patterns. Vertical lines show the position of nuclear Bragg peaks. The (*hkl*) values of the observed peaks are also listed. The additional magnetic broad peak is marked with asterisk. (b) The experimentally observed selected region of neutron powder diffraction patterns for all $YBaCo_{4-x}Fe_xO_7$ ($x = 0, 0.2, 0.4, 0.5, 0.6,$ and $0.8$) compounds at 22 K showing the evolution of the magnetic peak with the Fe-substitution. (c)-(e) Magnetic diffraction patterns at 22 K (after subtraction of the nuclear background at 130 K) for the $x = 0$, $x = 0.5$, and $x = 0.8$ compounds, respectively. The solid lines are the Lorentzian fits. (f) The obtained Lorentzian peak widths as a function of the Fe-concentration ($x$).